\documentclass[conference,doublecolumn]{IEEEtran}
\def\BibTeX{{\rm B\kern-.05em{\sc i\kern-.025em b}\kern-.08em
    T\kern-.1667em\lower.7ex\hbox{E}\kern-.125emX}}
\usepackage{cite}
\usepackage{caption}
\usepackage{subcaption}
\usepackage{graphicx}

\usepackage{amsmath}
\usepackage{amssymb}
\usepackage{amsthm}

\usepackage{stfloats}
\usepackage{bm}
\usepackage{xcolor}

\usepackage{tikz}
\usetikzlibrary{calc,shapes,arrows,positioning,plotmarks,shadows,calc,matrix,fit,patterns}
\usepackage{pgfplots}
\usepackage{todonotes}
\usepackage{datetime}
\usepackage{cancel}
\usepackage{hhline}
\usepackage{cases}
\usepackage{nicefrac}
\usepackage{etoolbox}
\usepackage{algorithm}
\usepackage{algorithmic}
\usepackage{cleveref}
\usetikzlibrary{plotmarks}
\usetikzlibrary{arrows.meta}
\usepackage{grffile}
\usepackage{dsfont}

\newcommand{\figref}[1]{Fig.~\ref{#1}}

\newtheoremstyle{remarkmod}
  {\topsep}   
  {\topsep}   
  {\normalfont}  
  {0pt}       
  {\itshape} 
  {.}         
  {5pt plus 1pt minus 1pt} 
  {}          
\theoremstyle{remarkmod}

\makeatletter
\newcommand{\ALC@comblock}[1]{\ifthenelse{\equal{#1}{default}}%
{}{\textbf{#1}}}
\newenvironment{ALC@bl}{\begin{ALC@g}}{\end{ALC@g}}
\newcommand{\BLOCK}[2][default]{
	\ALC@it\ALC@comblock{#1}\ #2\begin{ALC@bl}
}
\newcommand{\ENDBLOCK}{
	\end{ALC@bl}
}

\makeatother
\usepackage{graphicx}
\usepackage{mathtools, cuted}
\usepackage{siunitx}
\IEEEoverridecommandlockouts
\begin{document}
\title{{Optimized Frequency-Diverse Movable Antenna Arrays for Directional Secrecy in Wireless Systems} 	
\thanks{This work was supported by the German Federal Ministry of Research, Technology and Space (BMFTR) project 6G-ANNA [grant agreement number 16KISK095].}
}
\author{\IEEEauthorblockN{Chu Li \IEEEauthorrefmark{1} , Marjan Boloori  \IEEEauthorrefmark{1}, Eduard Jorswieck \IEEEauthorrefmark{2}, Aydin Sezgin    \IEEEauthorrefmark{1}
	\IEEEauthorblockA{\IEEEauthorrefmark{1} Ruhr-Universit\"at Bochum, Germany } 
	\IEEEauthorblockA{\IEEEauthorrefmark{2} Technische Universit\"at  Braunschweig, Germany }
		Email:  \{chu.li, marjan.boloori, aydin.sezgin\}@rub.de, e.jorswieck@tu-braunschweig.de}
}
\maketitle

	\begin{abstract}
Movable-antenna (MA) arrays are envisioned as a promising technique for enhancing secrecy performance in wireless communications by leveraging additional spatial degrees of freedom. However, when the eavesdropper is located in the same direction as the legitimate user, particularly in mmWave/THz bands where line-of-sight (LOS) propagation dominates, the secrecy performance of MA arrays becomes significantly limited, {thus directionally insecure}. To address this challenge, we employ a joint design that combines an MA array with a frequency-diverse array (FDA) at the transmitter {to secure the transmission across both range and direction. Specifically, we derive closed-form expressions for the optimal antenna positions and frequency shifts, assuming small perturbations in both parameters from a linear frequency-diverse MA configuration.} Furthermore, we compare the worst-case secrecy rate under this minor perturbation assumption with that obtained under a general constraint, where simulated annealing is employed to numerically determine the optimal parameters. Simulation results confirm that the proposed optimized frequency diverse MA approach significantly enhances secrecy performance in the presence of an eavesdropper aligned with the direction of the legitimate receiver.
\end{abstract}

\begin{IEEEkeywords}
Movable-antenna (MA), frequency diverse array (FDA), worst-case secrecy
\end{IEEEkeywords}
\section{Introduction}
Security is one of the primary concerns in future 6G networks, particularly for critical applications such as remote surgery, vehicle-to-everything (V2X) communications, and smart manufacturing, where highly sensitive and confidential information is transmitted. Traditionally, secure transmission is ensured through cryptographic mechanisms implemented at the upper layers. To complement these mechanisms, physical layer security (PLS) can be employed. A key focus of PLS is secure (confidential) transmission, with confidentiality typically evaluated in terms of secrecy performance. To enhance secrecy performance, techniques such as artificial noise generation, multi-antenna diversity, and relay-based strategies can be employed.

Movable-antenna (MA) arrays are proposed recently in \cite{zhu2023modeling}, where the positions of individual antennas can be dynamically adjusted within a defined region. By optimizing the antenna positions, the MA array can achieve improved secrecy performance compared to the conventional uniform phased array (CPA). {Beamforming and antenna positions are jointly optimized to enhance the secrecy rate in \cite{10416363} and to minimize the secrecy outage probability in \cite{10623758}. In these works, the antenna positions are determined using gradient descent. However, this method may converge to a local optimum due to the non-convex nature of the problem.} In \cite{cheng2024enabling}, a discrete sampling approach is adopted to determine the optimal antenna positions that maximize the secrecy rate. Specifically, the feasible region is discretized into grids, and a partial enumeration method with low complexity is proposed. Furthermore, transmit beamforming, artificial noise generation, and MA positioning are jointly optimized using block coordinate descent in a multiple-input multiple-output (MIMO) system in \cite{10684758}. 

However, the methods proposed in the aforementioned works are primarily effective when the eavesdropper is located in a direction different from that of the legitimate receiver. To address this challenge, frequency-diverse arrays (FDA) can be employed, wherein each antenna operates at a slightly different frequency. Originally introduced in radar applications \cite{1631800}, FDA has also attracted considerable attention in the field of PLS. By utilizing FDA with optimized frequency shifts at the transmitter, beamfocused transmission can be achieved, providing security in both angular and range dimensions. {More specifically, by formulating and solving various non-convex optimization problems, the frequency shifts of each antenna in the FDA are carefully designed to optimize the secrecy rate} \cite{8081593, 8078202, 10843324}. As an alternative, random FDA has been proposed to enhance secrecy \cite{hu2017artificial}. However, this approach may result in increased circuit complexity and cost at the transmitter.

In this work, we jointly employ MA and FDA at the transmitter to enhance the secrecy rate in the presence of multiple eavesdroppers. This combined configuration is referred to as a frequency-diverse MA. {In particular, we jointly design the antenna positions and frequency shifts of the frequency-diverse MA to address the directionally insecure scenarios encountered in mmWave/THz bands.} Initially, we consider general constraints for both antenna positions and frequency shifts, and apply an alternating optimization strategy using simulated annealing, which provides approximate globally optimal solutions. Additionally, we consider  small perturbations constraint in antenna positions and frequency shifts. In this context, analytical solutions are derived. Furthermore, we evaluate the performance of the proposed algorithms under critical eavesdropper placements: one eavesdropper is aligned in the same direction as the legitimate user, another is located at the same distance from the transmitter as the legitimate user, and a third is positioned within the main beam region of the linear FDA. Simulation results demonstrate that both approaches significantly enhance the secrecy rate and { provide directional secrecy}.


%

\section{System Model}
\label{sec:chmod}

	\begin{figure}[ht]
		\centering
	\resizebox{0.85\columnwidth}{!}{	
			\begin{tikzpicture}[scale=1.2, every node/.style={font=\small}, 
			user/.style={rectangle, draw, thick, minimum size=0.8cm, fill=white}]

			\begin{scope}[yshift=0.15cm]
				
				\draw[thick, rounded corners=1pt, line width=0.5pt] (-2.6,-0.15) rectangle (2.6,-0.5);
				\node at (0,-0.35) {\scriptsize Alice};
				
				\foreach \i/\x in {1/-2.1, 2/-1.05, M-1/1.05, M/2.1} {
					\draw[fill=black] (\x,0.0) -- ++(-0.1,0.2) -- ++(0.2,0) -- cycle; 
					\draw[line width=0.4pt] (\x,-0.15) -- (\x,0.0);                  
					\node at (\x,-0.85) {\scriptsize $x_{\i}$};
					\node at (\x,-0.65) {\scriptsize $f_{\i}$};
				}
				
			\end{scope}
			
			S
			\coordinate (bobpos) at (3.2,2.2);
			\node[user, fill=green!30] (bob) at (bobpos) {};
			\node at ($(bob)+(0.05,0.45)$) {Bob};
			\draw[->, thick] (0,0) -- (bob);
			\node at ($(bob)+(0.2, -0.7)$) {$(R_{B}, \theta_{B})$};
			
			\draw[fill=black] ($(bob.north west)+(0,0.2)$) -- ++(-0.15,0.3) -- ++(0.3,0) -- cycle;
			\draw[thick] (bob.north west) -- ($(bob.north west)+(0,0.2)$);

			\node[user, fill=red!20] (eve1) at (2.0,3.2) {};
			\node at ($(eve1)+(-0.7,0.45)$) {Eve$_2$};
			\draw[->, thick, dashed] (0,0) -- (eve1);
			\node at (1,3) {$(R_{E_2}, \theta_{E_2})$};
			\draw[fill=black] ($(eve1.north west)+(0,0.2)$) -- ++(-0.15,0.3) -- ++(0.3,0) -- cycle;
			\draw[thick] (eve1.north west) -- ($(eve1.north west)+(0,0.2)$);

			\coordinate (eve2pos) at ($(0,0)!1.5!(bobpos)$);
			\node[user, fill=red!20] (eve2) at (eve2pos) {};
			\node at ($(eve2)+(-0.7,0.45)$) {Eve$_1$};
			\draw[->, thick, dashed] (0,0) -- (eve2);
			\node at ($(eve2)+(0.2, -0.7)$) {$(R_{E_1}, \theta_{E_1})$};
			\draw[fill=black] ($(eve2.north west)+(0,0.2)$) -- ++(-0.15,0.3) -- ++(0.3,0) -- cycle;
			\draw[thick] (eve2.north west) -- ($(eve2.north west)+(0,0.2)$);

			\coordinate (eve3pos) at (-2.5,2.5);
			\node[user, fill=red!20] (eve3) at (eve3pos) {};
			\node at ($(eve3)+(-0.7,0.45)$) {Eve$_K$};
			\draw[->, thick, dashed] (0,0) -- (eve3);
			\node at (-2.7,1.7) {$(R_{E_K}, \theta_{E_K})$};
			\draw[fill=black] ($(eve3.north west)+(0,0.2)$) -- ++(-0.15,0.3) -- ++(0.3,0) -- cycle;
			\draw[thick] (eve3.north west) -- ($(eve3.north west)+(0,0.2)$);

			\draw[->] (0,0) -- (4.5,0) node[right] {X};
			\draw[->] (0,0) -- (0,4) node[above] {Y};
			
		\end{tikzpicture}
	}
				\caption{Frequency diverse MA assisted transmission in the presence of one legitimate user and multiple eavesdroppers}
				\label{fig:sys}
	\end{figure}
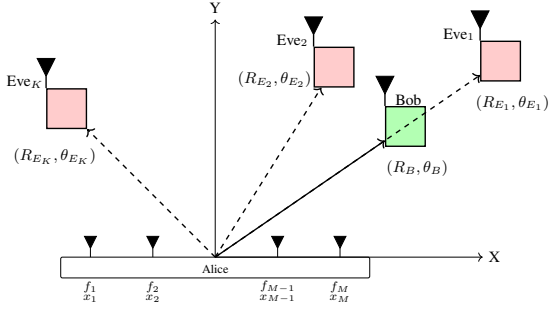                                                          

As shown in \figref{fig:sys}, we consider a scenario where a transmitter, Alice, equipped with $M$ movable antennas communicates with a legitimate receiver, Bob, in the presence of $K$ eavesdroppers, where $K<M$.
Both Bob and eavesdroppers are equipped with a single antenna\footnote{ {Note that in this work, we assume a single antenna at each of multiple eavesdroppers. However, the results remain the same if we instead assume a single eavesdropper equipped with $K$ antennas employing maximum ratio combining.} }. We use $\mathbf{x} = [x_1,\ldots,x_m,\ldots,x_M]^T$ to denote the antenna positions. Further, to ensure a direction-range secure transmission, we employ FDA at Alice, where each antenna operates at a slightly different center frequency. Specifically, the frequency at the $m$-th antenna is given by
\begin{align}
	\label{eq:frequency}
	f_m = f_0+\Delta F_m, \quad m \in {1,\ldots,M},
\end{align} 
where $f_0$ is the reference center frequency and $\Delta F_m$ is the frequency shift applied at $m$-th antenna. 
In particular, we have $\max_m(\Delta F_m)< 10^{-3} f_0$.

In this work, we investigate mmWave/THz transmission, where the LOS path dominates due to the severe attenuation experienced by NLOS paths, caused by high spreading, reflection, and scattering losses. Further, we  define the position of each receiver by a tuple $\boldsymbol{\psi}_u = (R_{\mathrm{u}}, \theta_{\mathrm{u}})$, where $R_{\mathrm{u}}$ is the distance from Alice to the receiver $u$, and $\theta_{\mathrm{u}}$ is the angle of arrival (AOA) at the receiver $u$. Here, we have $u \in \left\lbrace {\text{B}, \text{E}_1, \ldots, \text{E}_K}\right\rbrace $.  In this context, the channel between Alice and receiver $u$ can be modeled as
\begin{align}
	\label{eq:channel}
	\mathbf{h} (\mathbf{x}, \mathbf{f}, \theta_{\mathrm{u}}) = \sqrt{L_{\mathrm{U}}} \mathbf{a}(\mathbf{x},\mathbf{f}, \boldsymbol{\psi}_u),
\end{align} 	
where $L_{\mathrm{u}}$ represents the path loss factor. $\mathbf{f} \in \mathbb{C}^{M\times1} $ is the frequency vector, with its  $m$-th element $f_m$ given in \eqref{eq:frequency}, and $\mathbf{a}(\mathbf{x},\mathbf{f}, \boldsymbol{\psi}_u) \in \mathbb{C}^{M\times1}$ is the steering vector, whose $m$-th element is
\begin{align}
	a_m = 	e^{-j 2 \pi \frac{f_m}{c}\left( R_\mathrm{u}- x_m \cos\theta_\mathrm{u}\right)},
\end{align}	
where $c$ is the speed of light. Thereby, the received signal can be expressed as 
\begin{align}
	\label{eq:received_signal}
	y_{\mathrm{u}} = \sqrt{P}  \mathbf{h}(\mathbf{x}, \mathbf{f}, \boldsymbol{\psi}_u)^H \mathbf{w}x +n_{\mathrm{u}},
\end{align}
where $P$ is the transmit power, $\mathbf{w} \in \mathbb{C}^{M\times1}$ represents the beamforming vector with $\left\| \mathbf{w}\right\| ^2 =1$, and $n_{\mathrm{u}}$ is the receiver noise following a complex Gaussian distribution with zero mean and variance $\sigma_{\mathrm{u}}^2$.

Let $\gamma_\mathrm{u}$ denote the received signal-to-noise ratio (SNR) at the receiver $u$, which is calculated as
\begin{align}
	\gamma_\mathrm{u} = P \left| \mathbf{h}(\mathbf{x}, \mathbf{f}, \boldsymbol{\psi}_u)^H \mathbf{w}\right|^2 .
\end{align}
{According to \cite{zheng2022physical}, the worst-case secrecy rate, assuming that all eavesdroppers collaborate to jointly decode their received signals, is computed as 
\begin{align}
	\label{eq:sec_cap}	
	C_{\mathrm{E}_k}=  \left[ \log_2 (1+\gamma_\mathrm{B})-\log_2 \left( 1+\sum_{k=1}^{K}\gamma_{\mathrm{E}_k}\right) \right] ^+,
\end{align}  }
where $[x]^+ = \max(0,x)$.
%


\section{proposed algorithm}
\label{sec:Secrecy Analysis}
In this section, we first design the beamformer to maximize the received SNR at Bob. Subsequently, we optimize the antenna positions and frequency shifts alternately to minimize the received SNR at the eavesdroppers, under two different constraint settings. First, we consider general constraints on antenna spacing and frequency shifts. Second, we assume only small perturbations in antenna positions and frequency shifts from a conventional linear FDA, where both antenna positions and frequency shifts increase linearly {with antenna index.} By maximizing the SNR at Bob and subsequently minimizing the SNR at the eavesdroppers, the secrecy rate is effectively optimized. 
%
\subsection{Beamformer design}
By substituting \eqref{eq:channel} and \eqref{eq:received_signal} into \eqref{eq:sec_cap}, the received SNR at Bob is computed as
\begin{align}
	\label{eq:SNR_Bob}
	\gamma_\mathrm{B} &= \frac{P}{\sigma_\mathrm{B}^2} \left|  \mathbf{h} (\mathbf{x}, \mathbf{f}, \boldsymbol{\psi}_\mathrm{B})^H \mathbf{w}\right|^2 \nonumber \\
	& =  \frac{P}{\sigma_\mathrm{B}^2} L_\mathrm{B}  \left| \mathbf{a} (\mathbf{x}, \mathbf{f}, \boldsymbol{\psi}_\mathrm{B})^H \mathbf{w}\right|^2.  
\end{align} 
Maximizing \eqref{eq:SNR_Bob} is well-established, with the optimal solution given by the maximum ratio transmission (MRT) strategy, written as
\begin{align}
	\label{eq:beamforming}
\mathbf{w}^* =  \frac{1}{\sqrt{M}}\mathbf{a} (\mathbf{x}, \mathbf{f}, \boldsymbol{\psi}_\mathrm{B}).	
\end{align} 
As a result, we observe
\begin{align}
	\label{eq:SNR_Bob_2}
	\gamma_\mathrm{B} =  \frac{P}{\sigma_\mathrm{B}^2} L_\mathrm{B}  M.  
\end{align} 
Meanwhile, the received SNR at the $k$-th eavesdropper using $\mathbf{w}^*$ is computed as
\begin{align}
	\gamma_{\mathrm{E}_k} &= \frac{P}{\sigma_{\mathrm{E}_k}^2} \left|  \mathbf{h} (\mathbf{x}, \mathbf{f}, \boldsymbol{\psi}_{\mathrm{E}_k})^H \mathbf{w}\right|^2 \nonumber \\
	& =  \frac{P}{\sigma_{\mathrm{E}_k}^2} L_{\mathrm{E}_k}  {\frac{1}{M}\left| \mathbf{a} (\mathbf{x}, \mathbf{f}, \boldsymbol{\psi}_{\mathrm{E}_k})^H \mathbf{a} (\mathbf{x}, \mathbf{f}, \boldsymbol{\psi}_{\mathrm{B}})\right|^2}. 
\end{align}
Furthermore, we define the beampattern with frequency diverse MA as
\begin{align}
	\label{eq:beampattern}
	\eta (\mathbf{x},\mathbf{f}, \boldsymbol{\psi})=  \mathbf{a} (\mathbf{x}, \mathbf{f}, \boldsymbol{\psi})^H \mathbf{a} (\mathbf{x}, \mathbf{f}, \boldsymbol{\psi}_{\mathrm{B}}) 
\end{align}
Subsequently, {the worst-case secrecy rate in \eqref{eq:sec_cap} becomes to }
\begin{align}
	\label{eq:w-c_s}
	C_s^{\text{W-C}} = &\log_2 \left( 1+ \frac{P}{\sigma_\mathrm{B}^2} L_\mathrm{B}   M\right) \nonumber \\
	& -  \log_2\left( 1+ \sum_{k=1}^{K}\frac{P}{\sigma_{\mathrm{E}_k}^2}  \frac{L_{\mathrm{E}_k}}{M} \left| \eta (\mathbf{x}, \mathbf{f}, \boldsymbol{\psi}_{\mathrm{E}_k})\right|^2 \right) .
\end{align}
Considering the second term, we now formulate the optimization problem of interest as follows,
\begin{subequations}
	\begin{align}
			\label{eq:problem1}
			\min_{\mathbf{x}, \Delta \mathbf{F}} \quad	&\log_2\left( 1+\sum_{k=1}^{K}\frac{P}{\sigma_{\mathrm{E}_k}^2}  \frac{L_{\mathrm{E}_k}}{M} \left| \eta (\mathbf{x}, \mathbf{f}, \boldsymbol{\psi}_{\mathrm{E}_k})\right|^2 \right)  \\
			\text{s.t.} \quad 
			\label{eq:cond_1}
			&  -D \leq x_1<x_2<...<x_M\leq D,  \\
			\label{eq:cond_2}
			&   x_m-x_{m-1} \ge  \Delta D_{\text{min}} , \forall m \in \{2, \cdots, M\}, \\
			\label{eq:cond_3}
			&   \Delta F_{\text{min}} \leq \Delta F_m \leq \Delta F_{\text{max}}  , \forall m \in \{1, \cdots, M\},
		\end{align}
\end{subequations}
where constraint \eqref{eq:cond_1} ensures the antenna positions lie within the feasible spatial region, \eqref{eq:cond_2} enforces a minimum spacing $\Delta D_{\mathrm{min}}$ (typically set to $\lambda/2$) to avoid mutual coupling, and \eqref{eq:cond_3} bounds the frequency shifts within a practical range. The optimization problem in \eqref{eq:problem1} is non-convex with respect to both $\mathbf{x}$ and $\Delta \mathbf{F}$. Here, $\Delta \mathbf{F} \in \mathbb{C}^{M \times 1}$ is the frequency shift vector, with its $m$-th element denoted as $\Delta F_m$ in \eqref{eq:frequency}. Moreover, these two variables are coupled within \eqref{eq:problem1}, which further contributes to the non-convexity of the problem. Next, we address the problem under two different constraint settings. First, we consider general constraints on antenna spacing and frequency increments, as defined in \eqref{eq:cond_1}, \eqref{eq:cond_2}, and \eqref{eq:cond_3}. Alternatively, we consider the case where only small deviations in antenna positions and frequency shifts from a uniformly spaced linear FDA are allowed.
%
\subsection{General constraints}
We now consider the general constraints on antenna positions and frequency increments. To address the resulting non-convex problem, we employ simulated annealing \cite{bertsimas1993simulated}, which enables us to approximate a global solution for $\mathbf{x}$ and $\Delta \mathbf{F}$ with relatively low computational complexity. To this end, we define the cost function as
\begin{align}
	J(\mathbf{x},\mathbf{f}) =  \sum_{k=1}^{K}\frac{P}{\sigma_{\mathrm{E}_k}^2}  \frac{L_{\mathrm{E}_k}}{M}\left| \eta (\mathbf{x},\mathbf{f}, \boldsymbol{\psi}_{\mathrm{E}_k})\right| ^2.
\end{align}

We first optimize the antenna positions $\mathbf{x}$ with fixed frequency offset $\Delta \mathbf{F}$. Rather than optimizing the positions directly, we optimize the inter-element spacing $\mathbf{d}$, whose $m$-th element is $d_m = x_{m+1}-x_m, \forall m \in [1,..,M]$ for all $m \in [1,..,M]$. This transformation allows us to bypass the ordering constraint in \eqref{eq:cond_1} and  \eqref{eq:cond_2},  and instead simply enforce bounds on the antenna spacings, i.e., $\Delta D_{\mathrm{min}} \leq d_m\leq  \Delta D_{\mathrm{max},t}$. Note that $D_{\mathrm{max},t}$ is adaptive. At each iteration, we set $D_{\mathrm{max},t} = 2D - \sum_{i \neq m} \mathbf{d}_t(i)$, representing the maximum feasible separation for the selected antenna $m$. Unlike a genetic algorithm, the complexity of simulated annealing does not increase exponentially with the number of antennas. This is because, at each iteration, only one element of $\mathbf{d}$ is updated. More specifically, the algorithm begins with initializing a feasible antenna separation vector $\mathbf{d}_0$. During each iteration, the antenna separation vector denoted by $\mathbf{d}_t$ is perturbed by changing the value of only one element of $\mathbf{d}_t$, with the new value randomly generated from a uniform distribution $U(\Delta D_{\mathrm{min}}, \Delta D_{\mathrm{max},t})$. The perturbed solution is accepted if it results in a reduced cost. Otherwise, it is accepted with a probability of $\exp{\left(-\frac{\Delta J}{T_t}\right)}$, where $\Delta J$ represents the cost difference between the current and previous iterations. Here, $T_t$ denotes the temperature at the $t$-th iteration, which gradually decreases according to a decay parameter $\alpha$. {At high temperatures, the algorithm accepts worse solutions with higher probability to escape local minima. As the temperature decreases, it becomes more conservative and eventually settles into the best solution found.}  The pseudo-code for simulated annealing to obtain the optimal $\mathbf{x}$ is provided below. {Note that, random(0,1) in the algorithm means uniformly selecting a random parameter between 0 and 1.}
\begin{algorithm}
	\caption{Simulated Annealing to obtain $\mathbf{x}$}\label{alg:cap}
	\begin{algorithmic}[1]
		\REQUIRE Positions of the eavesdroppers $\boldsymbol{\psi}_{\mathrm{E}}$, frequency vector $\mathbf{f}$,  initialization of $\mathbf{d}_0$, cost function $J(\mathbf{x},\mathbf{f})$,
		\FOR{$t=1$ to max. number of iterations}
		\STATE {$T_t = \alpha T_{t-1}$, $\alpha <1$  } 
		\BLOCK[Choose neighbor of $\mathbf{d}_{t-1}$]{}
		\STATE	 Random select an integer $m$, $1\leq m \leq M-1$ 
		\STATE   Calculate the maximum feasible antenna separation for the selected antenna $m$, i.e.,  $\Delta D _{\mathrm{max},t}$
		\STATE   Generate a uniformly distributed random value $d_m$ in the interval $(\Delta D _{\mathrm{min}}, \Delta D _{\mathrm{max},t})$ 	
		\STATE  Generate $\mathbf{d}^*$ by setting its $m$-th element equal to $d_m$, while the remaining elements are the same as  in $\mathbf{d}_{t-1}$
		\STATE Calculate $\mathbf{x}^*$ according to  $\mathbf{d}^*$	        	
		\ENDBLOCK
		\STATE  \textbf{Calculate} $\Delta J = J(\mathbf{x}^*,\mathbf{f}) - J(\mathbf{x}_{t-1},\mathbf{f} ) $
		\IF{$\Delta J < 0$ \textbf{or} $\exp{(-\frac{\Delta J }{T_t})}\geqslant \text{random}(0,1)$ }
		\STATE $\mathbf{d}_t =\mathbf{d}^*$, $\mathbf{x}_t =\mathbf{x}^*$
		\ELSE
		\STATE $\mathbf{d}_t =\mathbf{d}_{i-1}$, $\mathbf{x}_t =\mathbf{x}_{i-1}$
		\ENDIF
		\ENDFOR
		\RETURN $\mathbf{x}_t$
	\end{algorithmic}
\end{algorithm}

Next, we optimize the frequency shifts while keeping $\mathbf{x}$ fixed. The procedure closely follows Algorithm \ref{alg:cap}, with the only difference being that the frequency shift at the selected antenna is randomly generated within the range $\left[ \Delta F_{\text{min}}, \Delta F_{\text{max}} \right]$ in each iteration.  Therefore, the detailed steps are omitted here.c The two optimization steps are performed alternately until convergence {to a predetermined accuracy}.

\subsection{Minor perturbations in antenna positions and frequency shifts}
We now assume minor perturbations in both the frequency shifts and antenna positions relative to a linear FDA with uniformly spaced antennas. Given this assumption, closed-form expressions for the frequency shifts and antenna positions can be derived to minimize the received SNR at the eavesdroppers. Specifically, the position and frequency shift of the $m$-th antenna are expressed as
\begin{align}
		x_m &= \left( m - \frac{M+1}{2} \right) \Delta D + \Delta x_m, 
		\\
\Delta F_m &=  \left( m - \frac{M+1}{2} \right) \Delta F+ \Delta f_m, 
\end{align}
where $\Delta x_m$ and $\Delta f_m$ represent small deviations from the uniform antenna spacing $\Delta D$ and uniform frequency spacing $\Delta F $, respectively. In this context, the worst case secrecy rate is determined by $\Delta \mathbf{x} =  [\Delta x_1,...,\Delta x_M]^T$, $\Delta \mathbf{f} =  [\Delta f_1,...,\Delta f_M]^T$, and the position of the eavesdroppers denoted by $\boldsymbol{\psi}_{\mathrm{E}_k}$. 

Assuming that the frequency shift vector $\Delta \mathbf{f}$ is given and $\Delta x_m$ is sufficiently small compared to $ \left( m - \frac{M+1}{2} \right) \Delta D$, $\forall m \in [1,...,M]$, the beampattern in \eqref{eq:beampattern} can be approximated as 
\begin{align}
	\label{eq:ccprameter}
	&\eta(\mathbf{x}, \mathbf{f}, \boldsymbol{\psi}_{\mathrm{E}_k}) = \sum_{m=1}^{M} e^{-j 2 \pi \frac{f_m}{c}\left( R_\mathrm{B}- R_{\mathrm{E}_k}- x_m \left( \cos\theta_\mathrm{B}-\cos\theta_{\mathrm{E}_k}\right) \right)}  \nonumber \\
	&	\overset{(a)}{:=}  \sum_{m=1}^{M} e^{-j \frac{2 \pi f_0}{c} \left( (m-\frac{M+1}{2}) \Delta D + \Delta x_m\right)  (\cos \theta_{\mathrm{E}_k} -\cos \theta_{\mathrm{B}})} 
	\nonumber \\
	& \qquad \times e^{-j \frac{2 \pi \Delta F_m }{c} x_m (\cos \theta_{\mathrm{E}_k} -\cos \theta_{\mathrm{B}})}  e^{-j \frac{2 \pi }{c}  \Delta F_m  (R_\mathrm{B} -R_{\mathrm{E}_k})}
	\nonumber \\
	& \overset{(b)}{\approx}   \sum_{m=1}^{M}  e^{-j \frac{2 \pi }{c} \left( f_0 (m-\frac{M+1}{2})\Delta D  (\cos \theta_{\mathrm{E}_k} -\cos \theta_{\mathrm{B}}) + \Delta F_m  (R_\mathrm{B} -R_{\mathrm{E}_k}) \right) }  \nonumber \\
	& 	 \qquad \times \left(1- j  \frac{2 \pi f_0}{c} \Delta x_m   (\cos \theta_{\mathrm{E}_k} -\cos \theta_{\mathrm{B}})\right)  \nonumber \\
	& = \! \sum_{m=1}^{M}\! \! e^{-j	\phi_{m,{\mathrm{E}_k}}} \! - \! j \frac{2\pi f_0}{c} (\cos \theta_{\mathrm{E}_k} \! - \!\cos \theta_{\mathrm{B}})\!\sum_{m=1}^{M} \! \! \Delta x_m e^{\!-j	\phi_{m,{\mathrm{E}_k}}},
\end{align}
where $e^{-j \frac{2\pi f_0}{c} (R_\mathrm{B} -R_{\mathrm{E}_k})}$ is omitted in step (a) since it does not affect the value of $\left| \eta \right|^2 $ which determines the worst secrecy rate in \eqref{eq:w-c_s}. (b) follows from the first-order Taylor expansion, i.e., $e^{jx} \approx 1+jx$.
Additionally, the term $ e^{-j \frac{2 \pi \Delta F_m }{c} x_m (\cos \theta_{\mathrm{E}_k} -\cos \theta_{\mathrm{B}})} $ is neglected in step (b) as $\Delta F_m \ll f_0 $. Also, the phase term $\phi_{m,{\mathrm{E}_k}} $ in \eqref{eq:ccprameter} is given by 
\begin{align}
	\label{eq:phase1}
	\phi_{m,{\mathrm{E}_k}} &= \frac{2 \pi }{c} \Big( f_0 (m-\frac{M+1}{2})\Delta D  (\cos \theta_{\mathrm{E}_k} -\cos \theta_{\mathrm{B}}) \nonumber \\ & \qquad \qquad+
	\Delta F_m  (R_\mathrm{B} -R_{\mathrm{E}_k})\Big).
\end{align} The real part and imaginary part of \eqref{eq:ccprameter} are, respectively, given by
\begin{align}
	&\text{Re} \left( \eta(\mathbf{x}, \mathbf{f}, \boldsymbol{\psi}_{\mathrm{E}_k}) \right) = \sum_{m=1}^{M} \cos(\phi_{m,{\mathrm{E}_k}} ) \\
	& \qquad  - \frac{2\pi f_0}{c} (\cos \theta_{\mathrm{E}_k} -\cos \theta_{\mathrm{B}}) 	\sum_{m=1}^{M} \sin(\phi_{m,{\mathrm{E}_k}} ) \Delta x_m, \nonumber
\end{align}
\begin{align}
	&\text{Im} \left( \eta(\mathbf{x}, \mathbf{f}, \boldsymbol{\psi}_{\mathrm{E}_k}) \right)  =
	\sum_{m=1}^{M} \sin(\phi_{m,{\mathrm{E}_k}} ) \\
	&\qquad -\frac{2\pi f_0}{c} (\cos \theta_{\mathrm{E}_k} -\cos \theta_{\mathrm{B}}) 	\sum_{m=1}^{M} \cos(\phi_{m,{\mathrm{E}_k}} ) \Delta x_m \nonumber.
\end{align}
We then determine $\Delta \mathbf{x}$ to ensure that the received SNR at each eavesdropper is nulled, i.e., 
\begin{align}
	\label{eq:nullSNR}
\frac{P}{\sigma_{\mathrm{E}_k}^2} \frac{L_{\mathrm{E}_k}}{M}   \left| \eta (\mathbf{x}, \mathbf{f}, \boldsymbol{\psi}_{\mathrm{E}_k})\right|^2 =0,  \forall k \in [1,...,K],
\end{align} thereby optimizing the worst-case secrecy rate.
Note that the phase terms $\phi_1, \phi_2, .., \phi_M$  are symmetric, i.e., $\phi_m \approx - \phi_{M-m+1}$, which arises from the assumption of only small deviations from a linear FDA configuration. Due to this symmetry, when the real part of $\eta$ is zero, the imaginary part will also be zero. Consequently, \eqref{eq:nullSNR} can be considered as $K$ linear equations, forming a linear system defined by $\mathbf{A}_{1} \Delta \mathbf{x} = \mathbf{b}_1$, where $\mathbf{A}_1$ is a $K \times M$ matrix with
\begin{align}\mathbf{A}_{1} \left( k, m\right)  = \frac{2 \pi f_0}{c} (\cos \theta_{\mathrm{E}_k} -\cos \theta_{\mathrm{B}}) 	 \sin(\phi_{m,{\mathrm{E}_k}} ).
\end{align}
Furthermore, $\mathbf{b}_{1}$ is a $K$ dimensional vector with 
\begin{align}\mathbf{b}_{1} \left(k \right)  = \sum_{m=1}^{M} \cos(\phi_{m,{\mathrm{E}_k}} ).
\end{align} The solution to the linear system is given by
\begin{align}
	\label{eq:solution1}
	\Delta \mathbf{x} = (\mathbf{A}_1^H \mathbf{Q} \mathbf{A}_1 + \alpha_1 \mathbf{I})^{-1} \mathbf{A}_1^H\mathbf{Q}\mathbf{b}_1,
\end{align} 
where $\mathbf{Q}$ is a $K \times K$ diagonal matrix whose $k$-th diagonal element is $\mathbf{Q}_{k,k} = \frac{P L_{\mathrm{E}_k}}{M\sigma_{\mathrm{E}_1}^2}$. Furthermore, $\alpha_1$ is a penalty parameter which enforces the $L2$ norm of $\Delta \mathbf{x}$ to be small \cite[Chap. 6]{boyd2004convex}. 

Similarly, given the antenna positions, the corresponding minor frequency shifts $\Delta \mathbf{f}$ can be determined by 
\begin{align}
	\label{eq:solution2}
	\Delta \mathbf{f} = (\mathbf{A}_2^H \mathbf{Q} \mathbf{A}_2 + \alpha_2 \mathbf{I})^{-1} \mathbf{A}_2^H\mathbf{Q}\mathbf{b}_2,
\end{align}
where
\begin{align}\mathbf{A}_{2,k,m} = \frac{2 \pi}{c} (R_\mathrm{B} -R_{\mathrm{E}_k}) 	 \sin(\varphi_{m,{\mathrm{E}_k}} ),
\end{align}
and
\begin{align}\mathbf{b}_{2,k} = \sum_{m=1}^{M} \cos(\varphi_{m,{\mathrm{E}_k}} ).
\end{align}
Furthermore, we have
\begin{align}
	\varphi_{m,{\mathrm{E}_k}} &= \frac{2 \pi }{c} \Big( f_0 x_m (\cos \theta_{\mathrm{E}_k} -\cos \theta_{\mathrm{B}}) \nonumber \\
		&\qquad  + (m-\frac{M+1}{2})\Delta F  (R_\mathrm{B} -R_{\mathrm{E}_k})\Big).
\end{align}
The derivation follows similar steps to those used in deriving the optimal position perturbations and is therefore omitted for brevity.


By applying \eqref{eq:solution1} and \eqref{eq:solution2} alternately we obtain the optimal antenna position adjustments and minor frequency shifts.

\section{Simulation results}
\begin{figure}
	\centering
	\includegraphics[width=1.1\linewidth]{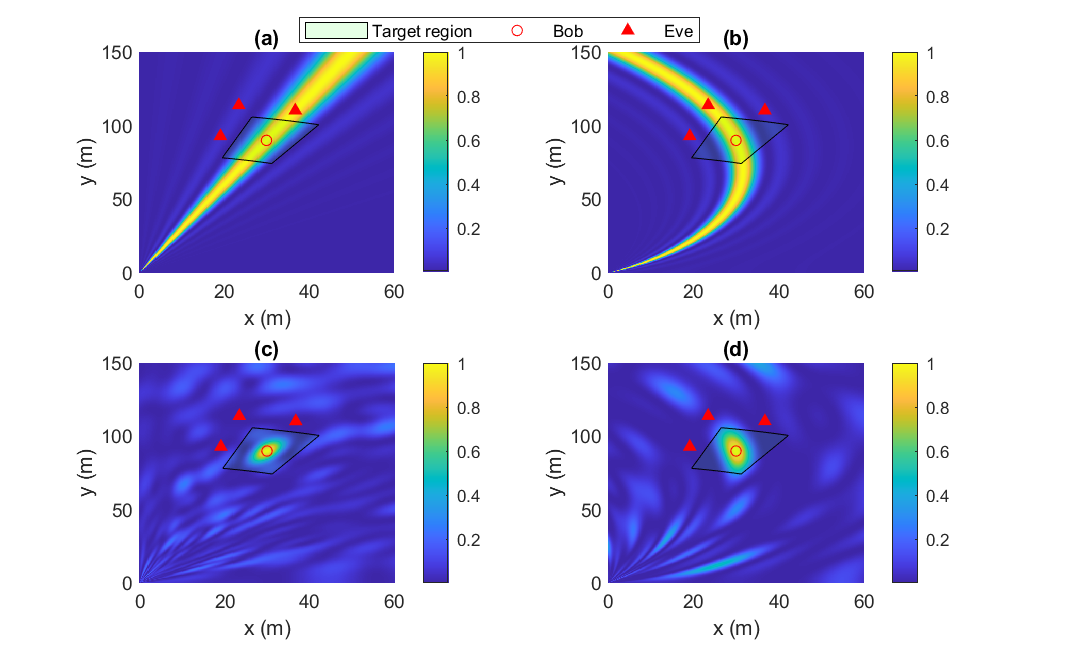}
	\caption{Normalized power of the beampattern with $M = 21$ and $K = 3$: (a) CPA, (b) linear FDA, (c) optimized FDMA (Opt. 1), (d) optimized FDMA (Opt. 2) }
	\label{fig:figure1}  \vspace{-1em}
\end{figure}

In this section, we present numerical results to demonstrate the effectiveness of the proposed algorithm. Throughout the simulations, we set the transmit power to $P = 5$ dBm, and the noise power to $\sigma_{\mathrm{u}}^2 = -80$ dBm, for all $u \in [{\text{B}, E_1, \ldots, E_K}]$. The path loss is modeled as $L_u (R_u) = L_0+ 25 \log_{10} (R_u) $ [dB], where $L_0 = 30$ dB  is the reference path loss. The center frequency is set to $f_0 = 30$ GHz. Furthermore, the center of the MA is placed at the origin, and the legitimate receiver Bob is located at $[ \SI{30}{\meter},  \SI{90}{\meter}]$. We compare the results with CPA, linear FDA, optimized frequency diverse MA under general constraint, referred to as FDMA (Opt. 1), and optimized frequency diverse MA under minor perturbations, referred to as FDMA (Opt. 2). Further, the optimization parameters are set as $\Delta F_\text{min} = -10$ MHz, $\Delta F_\text{max} = 10$ MHz, $\Delta D_\text{min} = 0.5 \lambda$ MHz, $\Delta F = -1$ MHz, $D = M \lambda $, and $\Delta D= 1.5 \Delta D_\text{min}$, where the choice $\Delta D= 1.5 \Delta D_\text{min}$ ensures the minimum antenna separation requirement ($0.5 \lambda$) is met under the minor perturbation assumption. 
More specifically, three critical eavesdroppers are included in the simulation (if not specified otherwise):
\begin{itemize}
	\item
    $\text{E}_1$ is located in the same direction as Bob, with a distance  $R_{\text{E}_1} =R_{\text{B}}  + \frac{3c}{2M \Delta F}$. This corresponds to the position where the maximum of the sidelobes occurs when using a linear FDA with frequency spacing $\Delta F$, under the condition $\theta_{\text{E}_1} = \theta_{\text{B}}$. 
	\item $\text{E}_2$ is placed at the  same distance from  Alice as Bob, and  $\theta_{\text{E}_2} = \arccos\left( \cos\left( \theta_{B}-\frac{3 \lambda}{2M\Delta D}\right) \right)$.  This corresponds to the angular location where the maximum of the sidelobes occurs when using a CPA with antenna separation $\Delta D$.
	\item $\text{E}_3$ is positioned within the main beam region if a linear FDA is used. Concretely, we use $\theta_{\text{E}_3} =  \theta_{\text{E}_2}$ and $R_{\text{E}_3} =R_{\text{E}_1}$.
\end{itemize}
	These cases represent the most challenging cases, as illustrated in \figref{fig:figure1}. In addition, we define the target region as
	\begin{align}
	\boldsymbol{\Psi}&= \bigg\lbrace  \left(  R, \theta \right) \left|R-R_B \right| \ge  \frac{c}{M \Delta F}, \nonumber \\
	&\left| \theta -\theta_{B} \right| \ge \arccos\left( \cos\left( \theta_{B}-\frac{\lambda}{M \Delta D}\right) \right)   \bigg\rbrace.
	\end{align}
	The boundary of the angular region is determined by the location of the first null in the beampattern $\eta$ when CPA is employed. Similarly, the boundary of the distant region is defined by the first null in $\eta$ when linear FDA is employed \cite{10843324}. The target region is shown as the area filled with transparent green in \figref{fig:figure1}.

\begin{figure}
	\centering
	\includegraphics[width=1\linewidth]{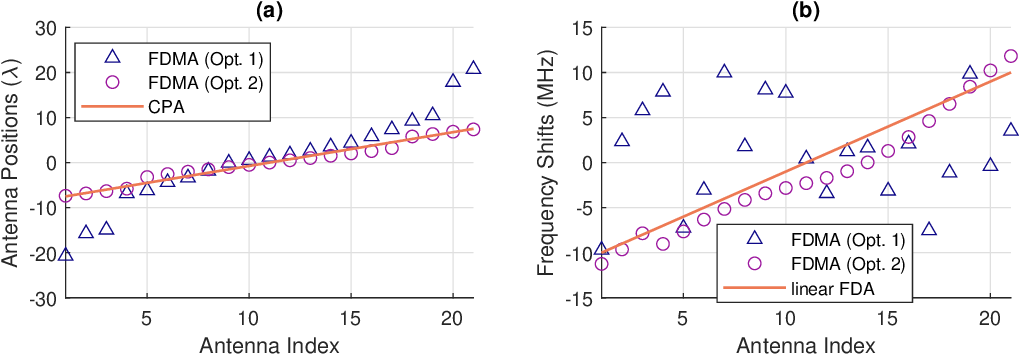}
	\caption{Comparison of the optimized antenna positions and frequency shifts using Opt. 1 and Opt.2}
	\label{fig:positionfreq} \vspace{-1em}
\end{figure}
In \figref{fig:figure1}, we illustrate  a map of the normalized beampattern power,  i.e., $ \frac{\left| \eta\right|^2}{M^2} $, observed at coordinates $[x, y]$. From \figref{fig:figure1} (a) and \figref{fig:figure1} (b), we observe maximum power within the main beam regions of the CPA and linear FDA configurations. An eavesdropper located within these regions can receive a high SNR, making it difficult to maintain the confidentiality of the transmitted information. In contrast, with both Opt. 1 and Opt. 2 (see \figref{fig:figure1}(c) and (d)), the maximum power is tightly focused in the target region, while the power is effectively suppressed at the locations of the eavesdroppers.
{Here, we observe that both approaches effectively address the issue of directional insecurity in the transmission.} 
 Additionally, \figref{fig:positionfreq}(a) compares the antenna positions scaled by $\lambda$ for FDMA (Opt. 1), FDMA (Opt. 2), and CPA. FDMA (Opt. 2) shows only minor deviations from the antenna positions of CPA, whereas FDMA (Opt. 1) exhibits more noticeable deviations. A similar trend is observed in \figref{fig:positionfreq}(b), which compares the corresponding frequency shifts (in MHz) across FDMA (Opt. 1), FDMA (Opt. 2), and linear FDA. 

\begin{figure}
	\centering
	\includegraphics[width=0.85\linewidth]{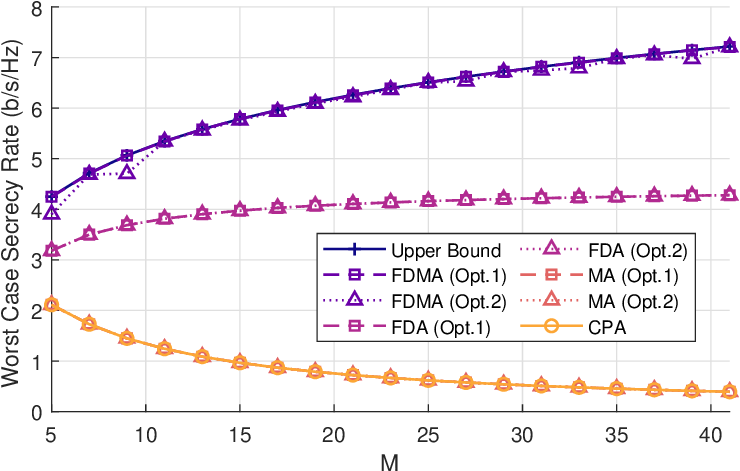}
	\caption{Worst case secrecy rate with respect to $M$ with $K = 3$}
	\label{fig:secrecyratevsmk3} \vspace{-1em}
\end{figure}

In \figref{fig:secrecyratevsmk3}, we compare the worst case secrecy rate under different configurations at Alice. The upper bound is determined by the rate at Bob without considering any eavesdropping. It can be observed that FDMA (Opt. 1) closely follows the upper bound and outperforms all other configurations. FDMA (Opt. 2) performs slightly worse than FDMA (Opt. 1) when $M$ is small, yet both configurations exhibit an increasing secrecy rate as the number of antennas $M$ increases. We also present results for the FDA configuration under large and minor perturbations, denoted as FDA (Opt. 1) and FDA (Opt. 2), respectively. In both cases, equally spaced antenna elements are used. The worst case secrecy rate achieved with the optimized FDA alone is lower than that achieved with the optimized FDMA. This is because the worst case secrecy rate is dominated by the SNR obtained by $E_2$, who is located at the same distance from Alice as Bob. In this scenario, frequency adjustment alone does not offer a secrecy advantage with respect to $E_2$. Furthermore, we evaluate the performance of the MA configuration under large and minor perturbations, denoted as MA (Opt. 1) and MA (Opt. 2). The secrecy rates for these setups, along with that of CPA, are identical and decrease with increasing $M$. This is because $\gamma_{\mathrm{E}_1}$ dominates the worst case secrecy in this scenario. The distance $R_{\text{E}_1} =R_{\text{B}}  + \frac{3c}{2M \Delta F}$ decreases with $M$, thus enhancing the channel gain at $E_1$. As a result, the worst case secrecy rate is decreasing with $M$. 

\begin{figure}[htbp]
	\centering
		\includegraphics[width=0.85\linewidth]{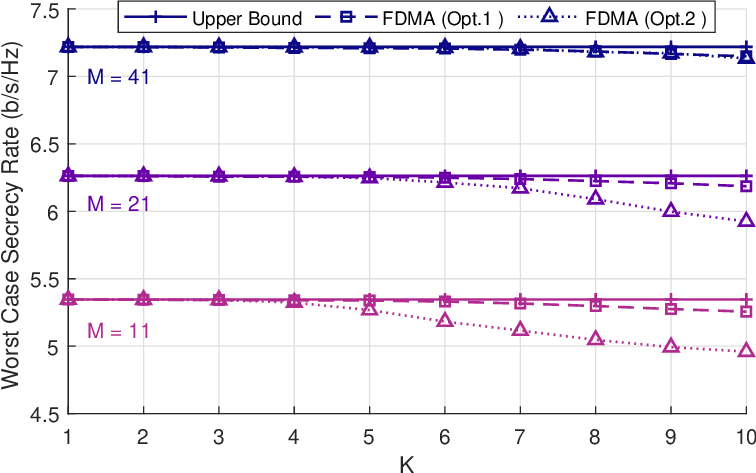}
	\caption{Worst-case secrecy rate with respect to $K$}
	\label{fig:main} \vspace{-1em}
\end{figure}
We now evaluate the performance of the proposed algorithms as a function of the number of eavesdroppers $K$. Specifically, the eavesdroppers are randomly positioned outside the  target region.
Fig. 4 shows the worst case secrecy rate as a function of  $K$. It can be observed that FDMA (Opt. 1) outperforms FDMA (Opt. 2) when $K$ is large. This is because FDMA (Opt. 1) allows for larger variations in both positions of the MA and frequency, thereby enabling greater design flexibility and adaptability compared to FDMA (Opt. 2). Additionally, as the number of eavesdroppers increases, the worst-case secrecy rate decreases for both algorithms due to the growing difficulty in simultaneously nulling multiple eavesdroppers.  However, with an increasing number of antennas $M$, the performance of both algorithms approaches the upper bound.

\section{Conclusion}
In this work, we investigated the worst case secrecy rate in a system employing a frequency diverse MA at Alice. The positions and frequency shifts of the antennas are jointly optimized to enhance secrecy performance. Two types of constraints were considered, a minor perturbation constraint and a more general constraint. For both cases, the antenna positions and frequency shifts were optimized accordingly. Simulation results demonstrate that under both constraint assumptions, a beamfocused transmission can be achieved, {providing directional secrecy}.
Furthermore, when the number of eavesdroppers is small, the performance of both optimization algorithms is nearly identical.
\label{sec:conclusion}

\bibliographystyle{IEEEbib} 
\bibliography{refs}  
\end{document}